\documentclass{article}
\usepackage{graphics}
\input epsf.sty
\begin{document}
\newcommand{\Abstract}[2]{{\footnotesize\begin{center}ABSTRACT\end{center}
\vspace{1mm}\par#1\par
\noindent
{~}{\it #2}}}

\newcommand{\TabCap}[2]{\begin{center}\parbox[t]{#1}{\begin{center}
 \small {\spaceskip 2pt plus 1pt minus 1pt T a b l e}
 \refstepcounter{table}\thetable \\[2mm]
 \footnotesize #2 \end{center}}\end{center}}

\newcommand{\TableSep}[2]{\begin{table}[p]\vspace{#1}
\TabCap{#2}\end{table}}

\newcommand{\FigCap}[1]{\footnotesize\par\noindent Fig.\ %
 \refstepcounter{figure}\thefigure. #1\par}

\newcommand{\TableFont}{\footnotesize}
\newcommand{\TableFontIt}{\ttit}
\newcommand{\SetTableFont}[1]{\renewcommand{\TableFont}{#1}}

\newcommand{\MakeTable}[4]{\begin{table}[htb]\TabCap{#2}{#3}
 \begin{center} \TableFont \begin{tabular}{#1} #4
 \end{tabular}\end{center}\end{table}}

\newcommand{\MakeTableSep}[4]{\begin{table}[p]\TabCap{#2}{#3}
 \begin{center} \TableFont \begin{tabular}{#1} #4
 \end{tabular}\end{center}\end{table}}

\newenvironment{references}%
{
\footnotesize \frenchspacing
\renewcommand{\thesection}{}
\renewcommand{\in}{{\rm in }}
\renewcommand{\AA}{Astron.\ Astrophys.}
\newcommand{\AAS}{Astron.~Astrophys.~Suppl.~Ser.}
\newcommand{\ApJ}{Astrophys.\ J.}
\newcommand{\ApJS}{Astrophys.\ J.~Suppl.~Ser.}   
\newcommand{\ApJL}{Astrophys.\ J.~Letters}
\newcommand{\AJ}{Astron.\ J.}
\newcommand{\IBVS}{IBVS}
\newcommand{\PASP}{P.A.S.P.}
\newcommand{\Acta}{Acta Astron.}
\newcommand{\MNRAS}{MNRAS}  
\renewcommand{\and}{{\rm and }}
\section{{\rm REFERENCES}}
\sloppy \hyphenpenalty10000
\begin{list}{}{\leftmargin1cm\listparindent-1cm
\itemindent\listparindent\parsep0pt\itemsep0pt}}%
{\end{list}\vspace{2mm}}
\def\TYLDA{~}
\newlength{\DW}
\settowidth{\DW}{0}   
\newcommand{\dw}{\hspace{\DW}}

\newcommand{\refitem}[5]{\item[]{#1} #2
\def\REFARG{#3}\ifx\REFARG\TYLDA\else, {\it#3}\fi
\def\REFARG{#4}\ifx\REFARG\TYLDA\else, {\bf#4}\fi
\def\REFARG{#5}\ifx\REFARG\TYLDA\else, {#5}\fi.}

\newcommand{\Section}[1]{\section{#1}}
\newcommand{\Subsection}[1]{\subsection{#1}}   
\newcommand{\Acknow}[1]{\par\vspace{5mm}{\bf Acknowledgements.} #1}
\newcommand{\Y}{Y_\ell^m}
\newcommand{\vxi}{\mbox{\boldmath{$\xi$}}}
\newcommand{\vx}{\mbox{\boldmath{$x$}}}
\newcommand{\vnab}{\mbox{\boldmath{$\nabla$}}}

\pagestyle{myheadings}

\def\thefootnote{\fnsymbol{footnote}}  
\begin{center}
{\Large\bf
On the distribution of the modulation frequencies of RR Lyrae stars\\ }
\vskip3pt {\bf Johanna Jurcsik$^{1}$, B\'ela Szeidl$^{1}$, Andrea Nagy$^{1}$, 
and \'Ad\'am S\'odor$^{1} $\\}
 \vskip3mm {$^1$
Konkoly Observatory of the Hungarian Academy of Sciences\\ 
P.O.~Box~67, H-1525 Budapest, Hungary\\
jurcsik,szeidl,nagya,sodor@konkoly.hu
}\vskip5mm
\end{center}

\abstract{For the first time connection between the pulsation and modulation 
properties of RR Lyrae stars has been detected. Based on the available data 
it is found that the possible range of the modulation frequencies, i.e, the 
possible maximum value of the modulation frequency depends on the pulsation 
frequency. Short period variables ($P<0.4$~d) can have modulation period as 
short as some days, while longer period variables ($P>0.6$~d) always exhibit 
modulation with $P_{mod}>20$ d. We interpret this tendency with the equality 
of the modulation period with the surface rotation period, 
because similar distribution of the rotational periods is expected if an upper 
limit of the total angular momentum of stars leaving the RGB exists. 
The distribution of the projected rotational velocities of red and blue 
horizontal branch stars at different temperatures shows a similar behaviour as
$v_{rot}$ derived for RR~Lyrae stars from their modulation periods. This common 
behaviour gives reason to identify the modulation period with the rotational period 
of the modulated RR~Lyrae stars.

Stars: variables: RR~Lyr --
            Stars: oscillations --
            Stars: horizontal-branch 
 }

\section{Introduction}

Different type modulations, which appear as close frequency component(s)
in the vicinity of the radial mode frequency in the Fourier spectrum of
RR Lyrae stars, seem to be a common property of both fundamental and first overtone
variables. The classical Blazhko (BL) phenomenon is just one special case when the
spectrum is characterized by equidistant triplets at the frequencies of the radial 
mode and its harmonics. Modulation manifests in doublet ($\nu_1$) and non-equidistant
triplet ($\nu_2$) structure of the Fourier spectrum, as well. 

The transition between the different type modulations  seems to be, however, 
continuous. The strong asymmetry of the amplitudes of the side frequencies of the
triplets as seen in many Blazhko stars, indicates that one of the side peaks may even
vanish to the noise level. In this case doublet structure of the spectrum is detected.
Though, according to our present knowledge, it cannot be excluded for sure that 
different physics govern the different types of modulation, a natural expectation is  
trying to find first a common explanation of the phenomenon. Nonradial modes triggered
by resonance effects or magnetic field seem to be the most promising solution for the
modulation. For a recent detailed summary of the problems of the theoretical 
explanations of the Blazhko effect see e.g., the Introduction in Dziembowski \& Mizerski
(2004).
 
All attempts to find any connection between the pulsation and modulation properties of
RR Lyrae stars have been yet failed. To find the correct answer of the modulation such 
a relation would be of great importance. Its benefit would be also to control the
reality of model predictions.

In this short note we have made a new attempt to find such a relation.
Our recent observations at Konkoly Observatory led to the unexpected discovery
of extreme short modulation periods of RR Geminorum and SS Cancri.
This fact directed our interest to the distribution of the modulation frequencies.

\section{The data}

\begin{figure}[h]
{\par\centering \resizebox*{10cm}{9.8cm}{\includegraphics{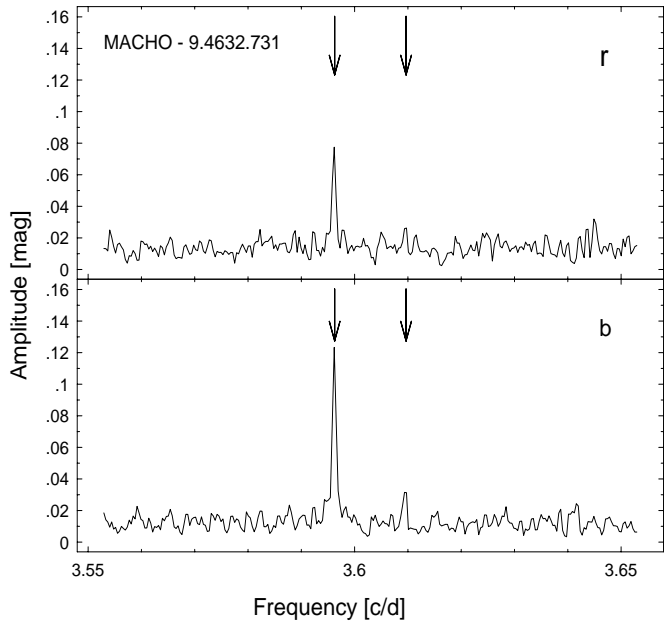}}
\par} \caption{{\small \label{jurcsikfig1}{\em 
Amplitude spectra of the prewhitened 'r' (top), and 'b' (bottom) data
of MACHO-9.4332.731, an RR1 star with $f1=3.60295$~c/d pulsation frequency.
This stars was classified as a non-modulated RR1 variable in Alcock et al. (2000).
The residual spectrum of the 8-year 'r' data shows one modulation peak at the
short frequency side of the removed pulsation frequency (i.e, doublet, $\nu_1$ spectrum),
while in the 'b' data modulation components both at the shorter and the longer
frequency sides (equdistant triplet, BL spectrum) are evident. 
}}\small }
\end{figure}

We have collected all the available data on the observed modulations of RR Lyrae stars. 
The large surveys (MACHO, OGLE) distinguish between the different types of modulation 
based on the structure of the Fourier spectrum of the stars. Variables exhibiting 
doublet structure in their spectra ($\nu_1$ variables) are most probably a special,
border-line case of those showing the equidistant triplet characteristics of the
classical Blazhko behaviour. The actually observed characteristics of the spectrum 
may be seriously affected by data length, distribution, S/N etc. For an example,
MACHO-9.4632.731 was classified as an RR1 variable in Alcock et. al (2000) from 6.5
years long 'r' data. According to the revision of the Alcock et. al (2000) results
utilizing the 8-year 'b' band MACHO data (Nagy \& Kov\'acs, 2005) the spectrum of this
star shows in fact asymmetric triplet structure. However, even in the 8 years of 'r'
data, because of its worse S/N properties, instead of a triplet seen in the 'b'
residual spectrum, only $\nu_1$, doublet structure can be resolved (see
Fig.~\ref{jurcsikfig1}).

\begin{figure}[h]
{\par\centering \resizebox*{9.8cm}{10.5cm}{\includegraphics{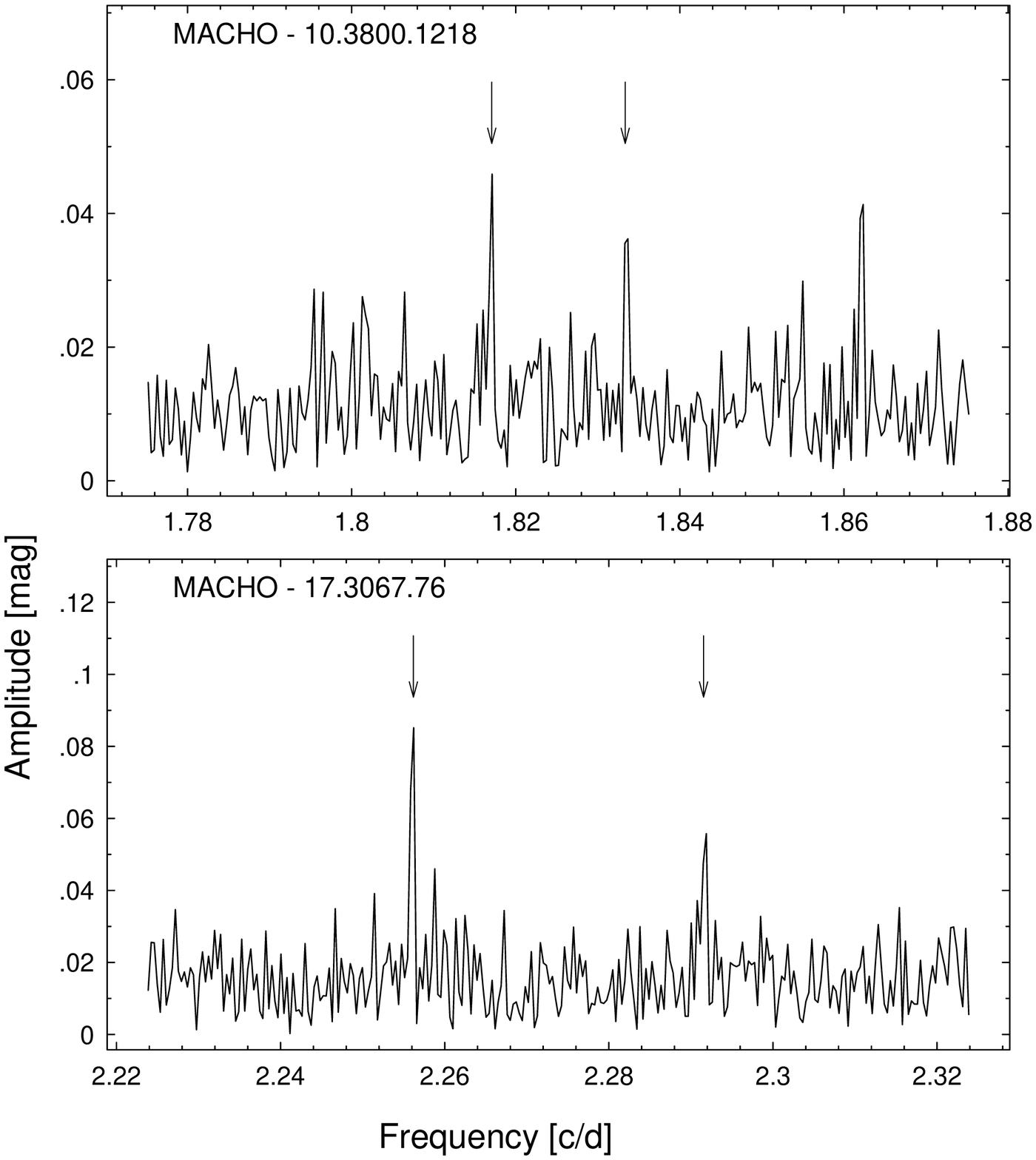}}
\par} \caption{{\small \label{jurcsikfig2}{\em 
Amplitude spectra of the residuals (after the removal of the fundamental mode 
pulsation) of two RR0 stars classified as BL1 ($\nu_1$) variables in Alcock et al. 
(2003) using MACHO $V$ data. MACHO-10.3800.1218 (top panel) shows in fact not only 
an equdistant triplet but also another modulation component with larger frequency
separation.
}}\small }
\end{figure}

We have also checked  the public MACHO database for hidden triplet structure of the
residual spectra of some of the fundamental mode $\nu_1$ variables classified in 
Alcock et. al (2003). Fig.~\ref{jurcsikfig2} shows two examples that these stars are
also intrinsically classical Blazhko variables showing triplets in their spectra.
The complete revision of the modulation properties of e.g. the MACHO RR Lyrae stars 
is, however, far beyond the scope of this note. We would only like to stress that 
somewhat better S/N statistics, more data, different treatment of filtering, etc.,  
can lead to a different classification of the modulation. Thus it can be supposed, 
that hidden triplet structure of the $\nu_1$ variables may probably be a common property.
In Jurcsik et al. (2005a) we have also shown a similar example. The extended data 
(N=3000) of RR Gem show symmetric triplets, while analyzing only 1000 datapoints, 
the residual spectrum becomes strongly asymmetric, indicating that with even a bit 
worse noise statistics, only one modulation component could have been detected. 
(For comparison, MACHO data are typically of $800-1000$ measurements / star.)

The modulation of the BL and $\nu_1$ variables can be characterized by a single
modulation period. As our interest is focused on the distribution of the modulation
periods, we restrict the data involved to these variables. Variables exhibiting
modulations with different periodicities ($\nu_2$ variables) are quite rare, and might
represent an indeed different type of modulation behaviour.  

The list of galactic field Blazhko stars (Smith, 1995, Table 5.2) has been completed
with new data given in Table~\ref{list}.
In some globular clusters (M5, M55, M68, NGC~6362, and $\omega$~Cen)
first overtone variables showing frequency doublets or triplets has been identified
(Olech et al., 1999a,b; Walker, 1994; Olech et al., 2001; Clement \& Rowe, 2000). 
V12 in M55 shows definite $\nu_2$ characteristics, thus it is not involved in this
study. V5 in M68 though clearly shows unstable light curve, its data have not been
analyzed by Walker (1994). Using those data we have found that the modulation period 
of this star is 7.45~d (frequency separation $\Delta f=0.134$ c/d).

Data of variables in the LMC, galactic bulge, and Sagittarius dwarf galaxy and its 
field were taken from Alcock et al. (2000), Alcock et al. (2003), Moskalik \& Poretti 
(2003), and Cseresnjes (2001), respectively.
The modulation frequency has been defined as the frequency separation of the radial
mode and the other close frequency appearing in the spectra. The sign of the separation,
i.e., that the modulation peak appears at shorter or longer frequency than that of  the
pulsation, has not been taken into account. 

Three objects have been eliminated from the sample. 
WY Dra has been removed from Smith's list of Blazhko stars, as we have found no convincing
evidence of its modulation either form the reanalyzis of Chis et al (1975)
$O-C$ and $M_{max}$ data, or according to the NSVS photometry (Wozniak et. al., 2004). 
We have found that MACHO 6.6212.1121 is not a modulated fundamental mode variable, 
but is a superposition of two RRab stars with 0.601390535~d and 0.5602746~d periods,
respectively. Its spectrum shows the harmonics of both components up to the 4., 5. order,
instead of showing modulation components at the same distance of one of the components and
its harmonics. It is a relatively bright RR Lyrae ($\overline{V}=19.059$~mag, the
magnitude range of the bulk of the LMC RRab stars is  $19.0 - 19.6$~mag) which supports
our solution. Another star excluded from the sample is vs4f114, a galactic field star in
the area of the Sagittarius dwarf galaxy. The period of this star (0.227~d) is anomalously
short for an RRc star. As its distance is not known, we suppose that it is rather a
multimode high amplitude $\delta$~Sct star. The omission of this object has, however, 
no effect on the results.

Alltogether data of 894 RR Lyrae stars which show single periodic modulation are utilized
(815 fundamental mode, 79 first overtone variables). The pulsation frequencies of the
first overtone variables were fundamentalized in order to obtain a homogeneous sample. 

\begin{table}[]
   \caption{Complementary list of galactic field Blazhko variables with known modulation
periods}
         \label{list}
\begin{tabular}{@{\hspace{0pt}}l@{\hspace{2pt}}c@{\hspace{2pt}}l@{\hspace{5pt}}l}
            \hline\hline
            \noalign{\smallskip}
Star& $P_{puls.}$ [d] & $P_{mod.}$ [d]& ref.\\
            \noalign{\smallskip}
            \hline
            \noalign{\smallskip}
ASAS\,81933-2358.2&0.2856 & \,\,\,\,\,\,8.1 & Antipin \& Jurcsik (2005)\\
SS Cnc       &0.3673 & \,\,\,\,\,\,5.3 & Jurcsik et al. (2005b)  \\
RR Gem       &0.3973 & \,\,\,\,\,\,7.2 & Jurcsik et al. (2005a) \\
V442 Her     &0.4421 & $>700$       & Schmidt \& Lee (2000) \\
GSC 6730-0109&0.4480 & \,\,\,26         & Wils \& Greaves (2004)   \\
AH Leo       &0.4663 & \,\,\,29         & Phillips \& Gay (2004) \\  
UX Tri       &0.4669 & \,\,\,43.7       & Achterbert \& Husar (2001) \\
FM Per       &0.4893 & 122              & Lee \& Schmidt (2001a) \\
GV And       &0.5281 & \,\,\,32         & Lee et al. (2002) \\
NSV 8170     &0.5510 & \,\,\,38.4       & Khruslov (2005)  \\
DR And       &0.5631 & \,\,\,57.5       & Lee \& Schmidt (2001b) \\
            \noalign{\smallskip}
            \hline
\end{tabular}
\end{table}

\section{Distribution of the shortest modulation frequencies}

\begin{figure}[h!!!!!]
{\par\centering \resizebox*{8cm}{8.7cm}{\includegraphics{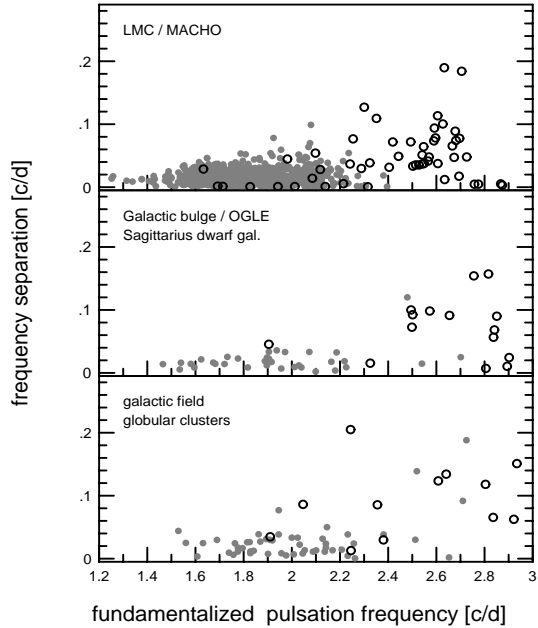}}
\par} \caption{{\small \label{jurcsikfig3}{\em The modulation frequencies 
(frequency separation) versus the pulsation frequency (fundamentalized for first
overtone variables) of the different samples of modulated RR Lyrae stars. 
Gray dots are for fundamental mode, while open circles denote first overtone 
variables. Each plot indicates clearly that short period modulation exists 
only in case of shorter period variables. The envelope of the data indicates that 
the shorter the period of the pulsation is the shorter the period of the modulation 
can be.  
}}\small }
\end{figure}

\begin{figure}[h]
{\par\centering \resizebox*{8cm}{7.8cm}{\includegraphics{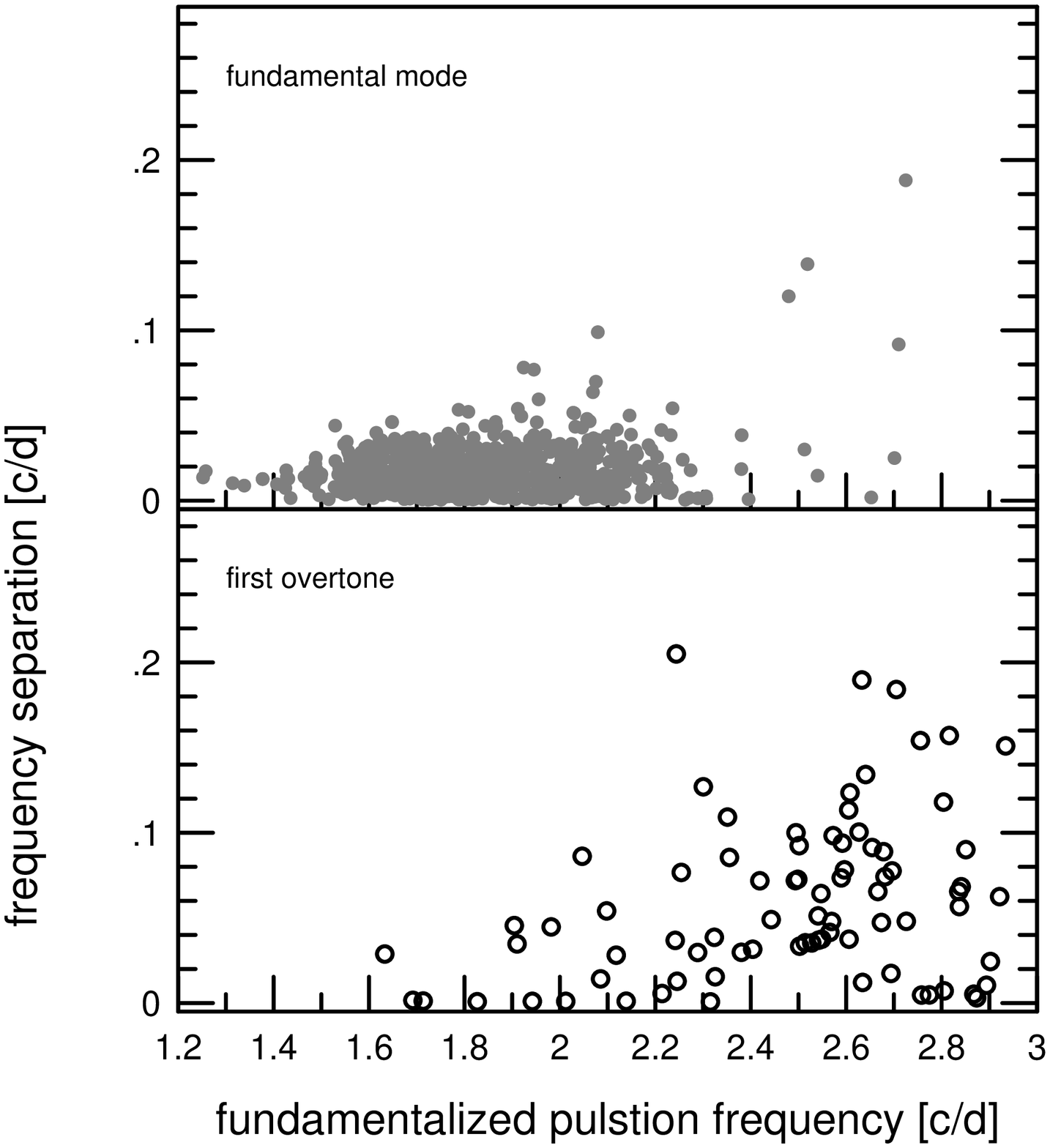}}
\par} \caption{{\small \label{jurcsikfig4}{\em The same as Fig~\ref{jurcsikfig3}, 
but the 815 fundamental mode and the 79 first overtone RR Lyrae stars are shown in
separate plots. Again, both plots show, that the frequency of the pulsation defines 
the possible range of the modulation frequencies.
}}\small }
\end{figure}

Investigating the distribution of the modulation frequencies at different
pulsation frequencies we have found that, instead of a uniform distribution, the
possible range of the modulation frequencies are wider towards larger pulsation
frequencies. The larger the pulsation frequency is, the larger the modulation 
frequency can be. This tendency does not mean that there would be a correlation 
between the two frequencies, as the modulation frequencies fill uniformly the 
$0- 0.04$~c/d range for the large majority of RRab stars with $1.6-2.2$~c/d
pulsation frequency, i.e, the typical modulation period is usually longer than
25~days. The shortest modulation period variables, however, do not follow this 
uniform distribution, instead, their modulation period is the shorter, the shorter
their pulsation period is. This tendency, that the shortest period modulation 
can be observed among the shortest period variables is equally evident in the 
different samples of the data, as shown in Fig.~\ref{jurcsikfig3}, and
Fig.~\ref{jurcsikfig4}. 

\begin{figure}[h]
{\par\centering \resizebox*{7.5cm}{8.4cm}{\includegraphics{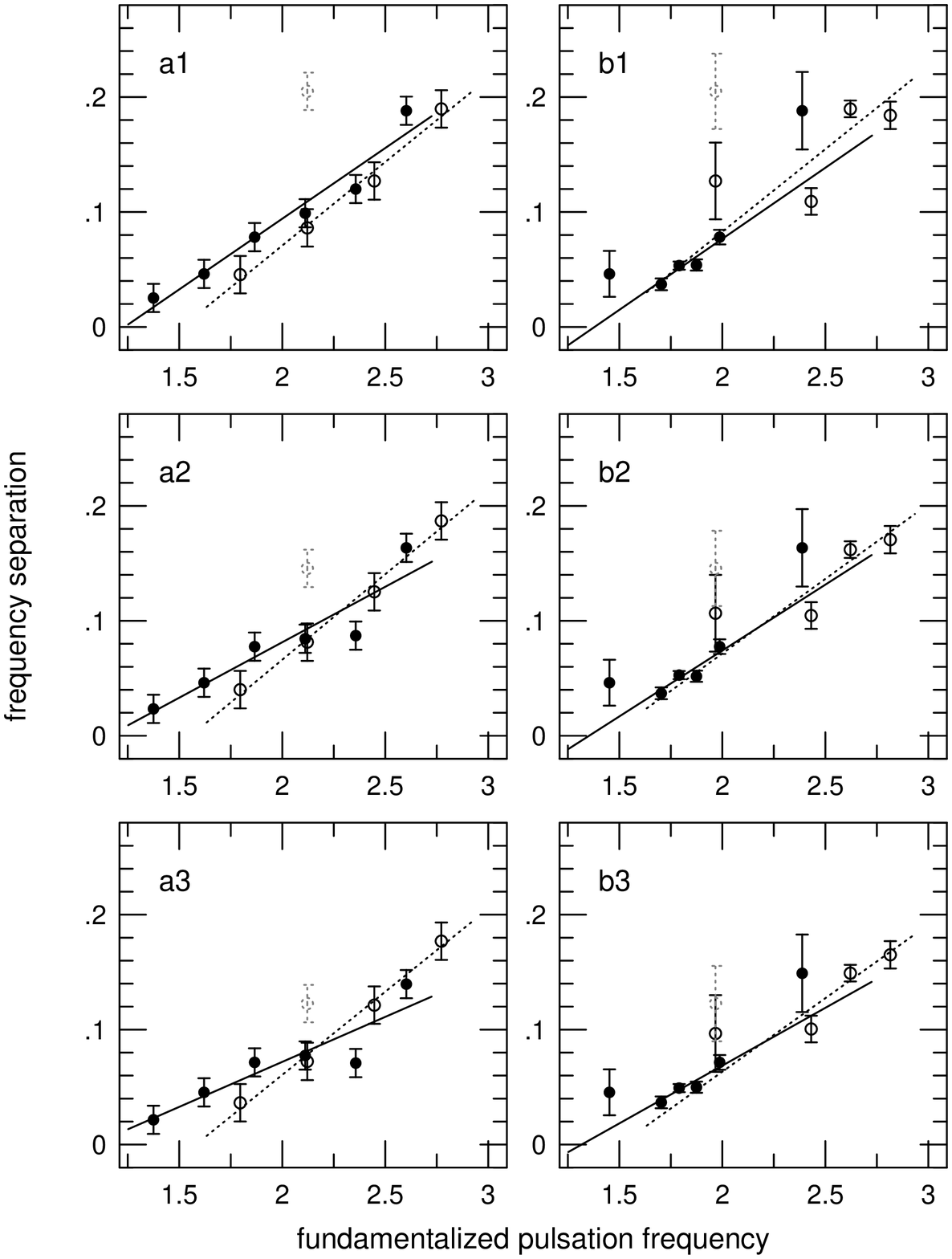}}  
\par} \caption{{\small \label{jurcsikfig5}{\em 
Distribution of the modulation frequencies of the 1, 2, and 3,  (from top to
bottom) variables with the shortest modulation periods.
Filled and open circles correspond to the RRab and the RRc samples shown in
Fig.~\ref{jurcsikfig4}, respectively. RRc sample including V104/M5 are shown 
by gray symbols. In the left panels bins are of equal widths (6 bins for RRab and 4
for RRc stars), while in 
the right panels of equal number of stars (136 for RRab, 20 for RRc stars).
Error bars indicate bin width, as because of the unknown intrinsic distribution
of the pulsation and modulation periods and the different biases affecting the 
samples used, a real estimate of the errors is impossible.
Weighted linear fits to  data (excluding V104/M5) are shown. In each plot the linear 
fits to the RRab 
and RRc data agree within the uncertainty limits, proving that the distribution of 
the shortest modulation period variables is the same for fundamental and first
overtone variables.
  }}\small }
\end{figure}

We have checked the modulation frequency distributions of RRab and RRc stars
separately. Data were divided into a) bins of equal width and b) bins of equal 
number of stars. The largest modulation frequencies, and the means of the 2, and 3 
 largest modulation frequencies in each bin are plotted in Fig.~\ref{jurcsikfig5}.
Error weighted linear fits to the data are also shown. V104 in M5 has been 
omitted from the fits, because as discussed in the next paragraph, there
may arise some 
doubt about its classification as a modulated RR Lyrae star. The fitted lines 
for the shortest modulation period RRab and RRc stars agree within the uncertainty
limits in each case, proving that the pulsation frequency dependence of the shortest
modulation periods is the same for fundamental and first overtone variables. 

\begin{figure}[h]
{\par\centering \resizebox*{8cm}{5.7cm}{\includegraphics{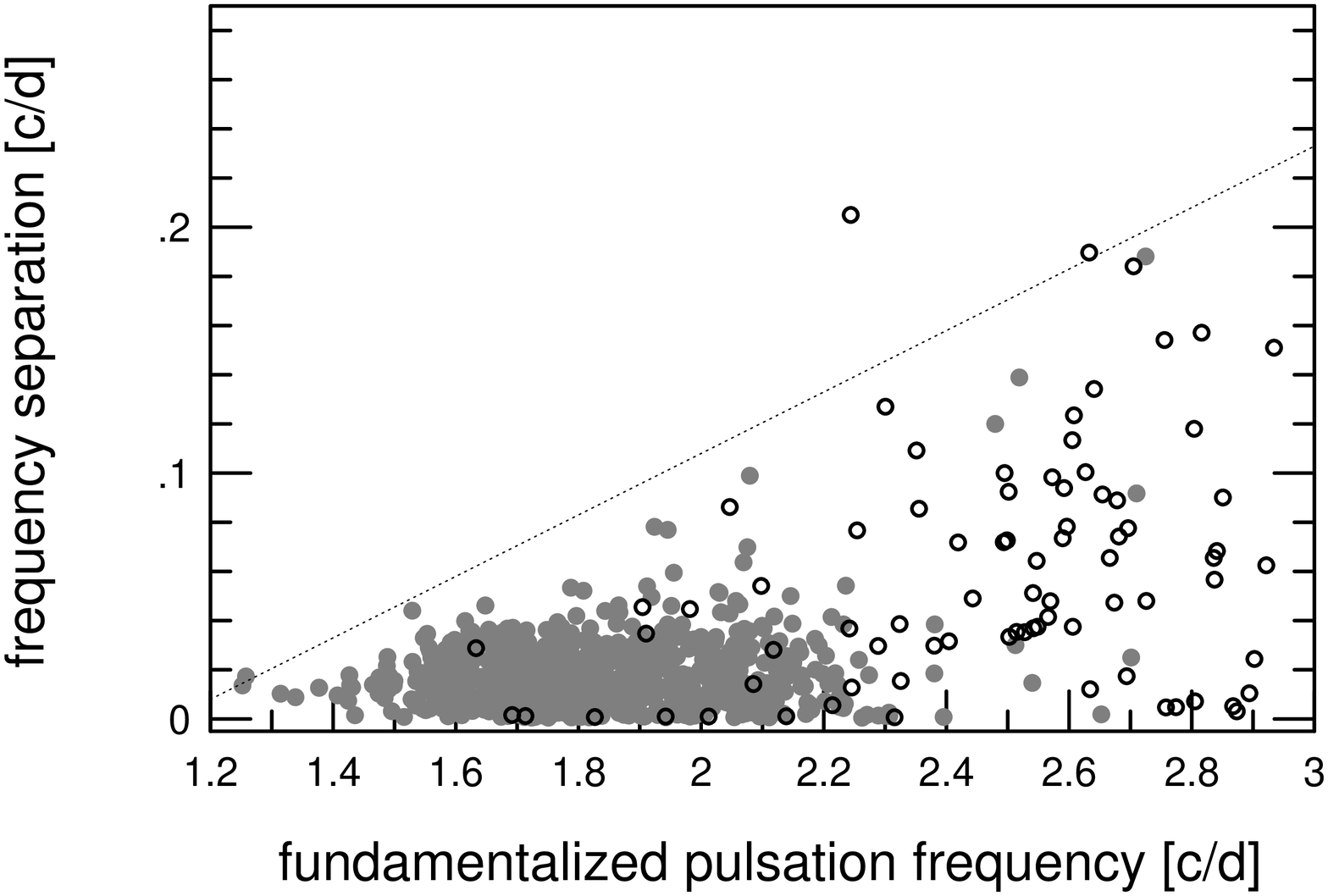}}
\par} \caption{{\small \label{jurcsikfig6}{\em 
The observed frequency separation versus the pulsation frequency for RR Lyrae stars
with known modulation frequencies are plotted. Gray dots denote fundamental, open
circles first overtone (with fundamentalized period) variables. Except one outlyer 
(V104/M5), the joint sample traces out a linear envelope for the possible modulation 
frequency as a function of the pulsation frequency.  
}}\small }
\end{figure}

The united sample of all the 894 RR Lyrae shown in Fig.~\ref{jurcsikfig6} defines
the limiting modulation frequency value as a linear function of the pulsation
frequency according to :  MAX~$(f_{mod})= 0.125 f_0 - 0.142 $. This relation is 
indicated by a dotted line in Fig.~\ref{jurcsikfig6}. There is  only one outlyer
with too large modulation frequency,  V104 in M5. This star has been already classified 
as a possible double mode variable (Reid, 1996), W~UMa type (Drissen \& Shara, 1998),
and as an RRc with two close frequencies (Olech et al., 1999b) allowing some suspect
that its frequency solution used is indeed correct.

The connection found between the possible range of the modulation frequencies 
and the pulsation frequency is the first direct link between the pulsation and
modulation properties of RR Lyrae stars. As any idea which try to explain the modulation 
connects it somehow to the rotation of the star, we suspect that the detected relation 
reflects also the rotational properties of the stars. In order to check this
possibility, we compared the data of RR Lyrae stars with projected rotational
velocity ($v\sin i$) observations of field blue and red horizontal branch stars (BHB,
RHB). To do so, we applied a crude transformation on the pulsation and modulation 
frequency data to get statistically valid temperature and rotational velocity values. 
The pulsation period was converted to temperature assuming that fundamental mode
variables with the shortest periods have $T_{eff}= 7500$~K temperature while the
longest period ones are at $T_{eff}= 5500$~K (i.e., $\log T_{eff}=3.97-0.289 P_0$).
The rotational velocity was derived from the $P_0\sqrt\rho=0.038$ pulsation
equation assuming uniform $M=0.55$ M$_{SUN}$  mass of the stars and that the observed
modulation frequency (frequency separation) corresponds to the rotational period. 
It can be easily seen that these approximations lead to a simple form of $v_{rot}$,
namely $v_{rot}= 362 f_{mod} \times f_0^{-2/3}$.

Projected rotational velocities of field BHB and RHB stars are taken from
Behr (2003b), Carney et al. (2003) and Kinman et al. (2000). For stars with multiple
$v\sin i$ measurements, the mean of the published values were considered.
Rotational properties of globular cluster BHB stars were studied, among others,  by
Behr (2003a), and Peterson et al. (1995). These are the only studies where both
temperature and $v\sin i$ data are available for BHB stars, therefore  globular
cluster data are taken from these sources.

\begin{figure}[h]
{\par\centering \resizebox*{9cm}{7cm}{\includegraphics{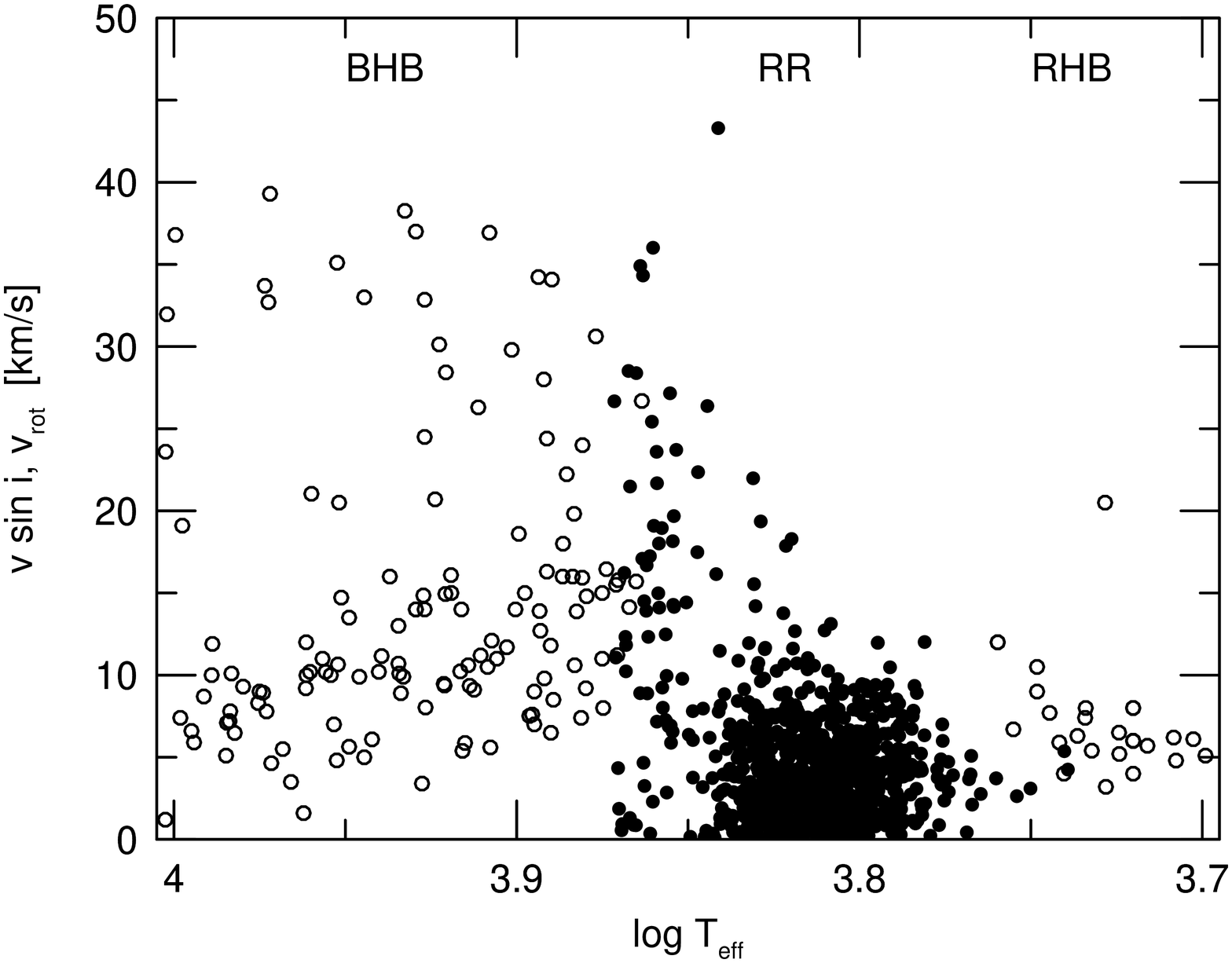}}  
\par} \caption{{\small \label{jurcsikfig7}{\em The observed rotational 
properties ($v\sin i$) of field blue and red horizontal branch stars (BHB, RHB -- open
circles) and rotational velocities of RR Lyrae stars (dots), -- assuming that the
detected modulation frequency corresponds to the surface rotational velocity of the 
star --, are plotted versus $\log T_{eff}$. The three samples follow the same trend, 
the hotter the stars are the higher their rotational velocity can be. There are one 
RHB star (HD 195636) and one globular cluster RR Lyrae star (V104/M5) with rotational 
velocity in excess to the global tendency defined by the tree samples.
The same trend of the projected rotational velocities of BHB and RHB stars as
seen from the modulation periods of RR Lyrae stars is the first direct evidence 
that the modulation period reflects, indeed, the rotational period of the stars.   
}}\small }
\end{figure}

Fig. 8 in Behr (2003b), and Fig.~5 in Recio-Blanco (2004) indicate already that the
maximum possible $v\sin i$ of both BHB and RHB stars show a temperature dependence, 
the hotter stars can rotate faster both in the $3.7 < \log T < 3.77$ RHB and in the
$3.87 < \log T < 4.00$ BHB temperature ranges.

The similar dependence of the possible largest rotational velocity values 
on the temperature as shown for BHB, RR Lyrae and RHB stars in Fig.~\ref{jurcsikfig7} 
gives a convincing proof of our assumption that the modulation periods of RR~Lyrae 
stars equal to their rotational period.

Behr (2003b) has already noticed the temperature dependence of the $v\sin i$
of RHB stars and connected it to a fixed value of the total angular momentum on
the horizontal branch. The combined sample of BHB, RR Lyrae and RHB stars 
points to that, there may be a common upper limit of the total angular
momentum of HB stars, most probably due to the mass/angular momentum loss of these 
stars during the termination of their RGB evolution.

\section{Concluding remarks}

In this note we have shown that the possible range of the modulation periods of
RR Lyrae stars depends on the pulsation period, i.e., on the physical 
properties of the stars. The hotter, smaller, short period ($P<0.45$~d) 
variables' modulation period can be as short as some days, while for 
the cooler, larger, long period ($P>0.6$~d) variables the modulation period 
cannot be shorter than $\sim 20$~d. This behaviour can be explained by a 
unique maximum value and a similar distribution of the angular momentum of 
the stars at different temperatures on the horizontal branch. The distribution 
of the projected rotational velocity measurements of RHB and BHB stars is a strong
confirmation of the validity of this idea.   

Though rotation has been connected to the modulation periods of RR Lyrae
stars already in many theoretical works, no direct evidence of rotation of 
RR Lyrae stars has yet been found. Peterson et al. (1996) gave an upper limit  
$v\sin i < 10$ km/s for each of the RR Lyrae stars they studied. The shortest 
known modulation period galactic RRab stars are SS Cnc, RR Gem, AH Cam, 
and Z CVn. The rotational velocities of AH Cam, and Z CVn are about 21 and 14~km/s
assuming 4.5, and 6.5 $ R_{SUN}$ radius, respectively (according to
their periods), and applying the simple formula 
$v_{rot} {\rm{[km/s]}} = 50 R {\rm{[R_{SUN}]}} \times f_{rot} {\rm{[1/d]}}$.
The rotational velocities of RR Gem and SS Cnc would be $ 30-40$  km/s.
For all these stars Peterson et al. (1996) found $v\sin i < 10$ km/s.
 
The similar behaviour of the rotational properties of RHB, BHB and RR Lyrae stars as
shown in Fig.~\ref{jurcsikfig7} is a direct evidence that the modulation period
reflects, indeed, the rotational period of RR Lyrae stars. We suggest that  the
contradiction between the predicted and observed values of the rotational velocities 
arises from projection effect, the modulation period corresponds to the true $v_{rot}$
of the stars, while $v\sin i$ is spectroscopically observed. The very small modulation
amplitudes of RR Gem and SS Cnc support this explanation. A circumspect 
spectroscopic analyses of the line broadenings of the modulated RR Lyrae stars would be
anyway of great importance in solving the mystery of the Blazhko phenomenon.

\Acknow
This research has made use of the SIMBAD database, operated at CDS Strasbourg,
France. The financial support of OTKA grants T-043504, T-046207 and T-048961 is
acknowledged. 
This paper utilizes public domain data obtained by the MACHO Project, 
jointly funded by the US Department of Energy through the University of 
California, Lawrence Livermore National Laboratory under contract No. 
W-7405-Eng-48, by the National Science Foundation through the Center for 
Particle Astrophysics of the University of California under cooperative 
agreement AST-8809616, and by the Mount Stromlo and Siding Spring 
Observatory, part of the Australian National University.
This publication makes use also of the data from the Northern Sky Variability
Survey created jointly by the Los Alamos National Laboratory and University of Michigan.
The NSVS was funded by the Department of Energy, the National Aeronautics and Space
Administration, and the National Science Foundation.

\end{document}